\documentclass[12pt, preprint, flushrt]{aastex}

\usepackage{graphicx}
\usepackage{natbib}
\usepackage{amssymb}
\newcommand{\refeqn}[1]{Eq. (\ref{#1})}

\begin{abstract}
Current plans call for the first Terrestrial Planet Finder mission, TPF-C, to be a monolithic space telescope with a coronagraph for achieving high contrast.  The coronagraph removes the diffracted starlight allowing the nearby planet to be detected.  In this paper, we present a model of the planet measurement and noise statistics.  We utilize this model to develop two planet detection algorithms, one based on matched filtering of the PSF and one using Bayesian techniques.  These models are used to formulate integration time estimates for a planet detection with desired small probabilities of false alarms and missed detections.
\end{abstract}


\begin{document}

\title{Linear and Bayesian Planet Detection Algorithms for the Terrestrial Planet Finder}
\author{N. Jeremy Kasdin}
\affil{ Mechanical and Aerospace Engineering,  Princeton University, Princeton, NJ  08544}
\email{jkasdin@princeton.edu} 
 \author{Isabelle Braems}
 \affil{ Laboratoire d'Etudes des MatŽriaux Hors-Equilibre (LEMHE),B‰t.413 , UniversitŽ Paris Sud, 91405 Orsay Cedes, TŽl : 01.69.15.46.77}
 \email{isabele.braems@lemhe.u-psud.fr }

\maketitle

\pagenumbering{arabic}

\section{Introduction}
\label{sec:intro}

In 2004, NASA  announced that it plans to launch the Terrestrial Planet Finder-C (TPF-C), a space based visible light telescope equipped with a coronagraph to detect earthlike planets around other stars, by 2016.  Work is proceeding on developing a baseline design that would meet the scientific requirements.  This currently consists of a $8 \times 3.5$ meter off-axis Cassegrain optical telescope combined with a starlight suppression system (SSS)  to achieve the needed starlight rejection.  The SSS is equipped to utilize either conventional Lyot coronagraphs or pupil masks (shaped or apodized).  At Princeton, we have been studying  what we call shaped pupil coronagraphs as an alternative to the traditional Lyot design. \citep{ref:KVSL,VSK02,VSK03,Kasdin04,Van04}  These are coronagraphs with apodized entrance pupils that rely solely on one/zero binary openings, resulting in a more manufacturable and robust design.

Critical to TPF-C's success is the ability to survey as many stars as possible and maximize the chances of detecting potential planets.  To do this, a thorough understanding of the data analysis is required in order to obtain reasonable estimates of the integration time for each observation.  \cite{Brown05} discusses at length the question of survey completeness and its relation to integration time.  In this paper, we focus on a more careful and complete model of the measurement and noise statistics and develop a probabilistic detection criteria.  We also analyze a Bayesian detection approach to see if detection time improvements can be made.  These expressions can be used for further completeness studies and estimates of the efficacy  of the TPF-C search program.

\section{Point Spread Function}
Before describing the detection approach, in this section we develop the needed definitions, units, and notation for photometry.  
We will assume a perfect imaging system and scalar Fraunhoffer theory such that the electric field at the image plane of the camera is the Fourier Transform of the field entering the camera at the exit pupil of the starlight suppression system (SSS),
\begin{equation}
\label{eq:fraunhoffer}
E_f(u,v) = \frac{e^{j \frac{k}{2f}(u^2+v^2)}}{j\lambda f} \int{\int_S{ E_o(x,y) e^{-j \frac{2\pi}{\lambda f}(xu + yv)} dx dy}}
\end{equation}
where $E_o$ is the electric field at the exit pupil aperture, $f$ is the focal length of the camera, $\lambda$ is the wavelength of the light, and the set $S$ represents the exit pupil aperture, which, for an open square aperture, as an example, is given by,
\[ S=\{(x,y):-D/2\le x \le D/2, -D/2\le y \le D/2\}  \]
where $D$ is the width of the aperture.

The optical system prior to the exit pupil of the SSS consists of various reflections and correction elements all of which have a certain efficiency.  If we take $E_i(\lambda)$ to be the uniform electric field  at the entrance pupil, then $E_o$ can be written,
\[ E_o(X,Y,\lambda) = \eta \mathcal{A}(X,Y) E_i(\lambda) \]
where $\eta < 1$ is the overall attenuation  of the optical elements up to the exit pupil (including any reduction due to an image plane mask), $\mathcal{A}(X,Y)$ is the apodization of the exit pupil, and we have explicitly shown the dependence on wavelength.  For a classical or bandlimited coronagraph, $\mathcal{A}(X,Y)$ is zero-one valued and just the Lyot stop  shape.  For a shaped pupil coronagraph, $\mathcal{A}(X,Y)$ represents the entire starlight suppression and is also zero-one valued.  

For a point source, the image plane detector measures the irradiance of the arriving field, $I_f=\frac{c
\epsilon_o}{2h}|E_f|^2$, in photon sec$^{-1}$ m$^{-2}$ $\mu$m$^{-1}$ where 
$c$ denotes the speed of light, $h$ is Planck's constant and $\epsilon_o$ is
the permittivity of free space. 
Squaring the magnitude of the electric field in
\refeqn{eq:fraunhoffer} gives the irradiance at the detector, 
\begin{eqnarray}
\label{eq:I}
I_f(\lambda,u,v) &=& \eta^2 \frac{c \epsilon_o}{2h} |E_i(\lambda)|^2 P(\lambda,u,v) \nonumber \\
& = &  \eta^2 I_i(\lambda) P(\lambda,u,v) 
\end{eqnarray}
where $I_i(\lambda) = \frac{c \epsilon_o}{2h}  |E_i(\lambda)|^2$ is the irradiance of the point source at the reference wavelength and $P(u,v)$ is the dimensionless point spread function (PSF) associated with the exit pupil of the SSS,
\begin{equation}
\label{eq:PSF}
P(\lambda,u,v) =\frac{1}{\lambda^2 f^2} \left | \int{\int_S{\mathcal{A}(x,y) e^{-j \frac{2\pi}{\lambda f}(xu + yv)} dx dy}} \right | ^2
\end{equation}
%
We note that the value of the integral in \refeqn{eq:PSF} at the origin is the square of the integral of the apodization, which for an open aperture (or shaped pupil or Lyot stop) is the square of the area of the exit pupil.  We use $A$ to represent the area of the entrance pupil of the telescope and $A_c$ to represent the area of the exit pupil of the coronagraph (Lyot stop or shaped pupil).   For an open telescope with no coronagraph, $A_c=A$.  For apodized pupils, we call $A_c$ the pseudo-area as it includes the reduction in throughput due to the continuous apodization.

There are three expected sources that produce electric field entering the telescope, the parent star ($I_s = \frac{c \epsilon_o}{2h}  |E_s(\lambda)|^2$), the planet ($I_p = \frac{c \epsilon_o}{2h}  |E_p(\lambda)|^2$) and the exo-zodiacal dust.  While the first two sources are point sources, the exo-zodi is a uniform extended source with flux $\mathcal{F}_{ez}(\lambda)$ in units of photon sec$^{-1}$ m$^{-2}$ $\mu$m$^{-1}$ ster$^{-1}$. 

For a planet offset on the sky by the angles $(\theta_x,\theta_y)$ the electric field at the entrance pupil becomes $E_p(\lambda) e^{2\pi i (\theta_x x/\lambda + \theta_y y/\lambda)}$ and by the Fourier Shift Theorem the irradiance at the detector becomes,
\begin{equation}
\label{eq:shifted_planet}
I_f(\lambda,u,v) = \eta^2 I_p(\lambda) P(\lambda,u-\theta_x f,v - \theta_y f) 
\end{equation}

For a diffraction limited system, we can assume that the uniform exozodical flux also creates a uniform flux density on the image plane and the  details of the PSF are unimportant (except for perhaps a slight broadening at the edges).  Thus, the irradiance at a single pixel, $(i,j)$, is given by $I_{ez_{ij}} = \mathcal{F}_{ez}(\lambda) \Omega_{ij} \Delta \alpha_{ij}$, where $\Omega_{ij}$ is the solid angle on the sky subtended by a single pixel and $\Delta \alpha_{ij}$ is the area of a pixel.  

\section{Measurements}
\label{sec:snr}


In this section we develop a model for the detector measurements and corresponding noise statistics.  This is essential for developing the detection algorithm and corresponding detection probabilities.  We assume here that the wavefront control system has successfully reduced the aberrated light to a uniform background in the discovery region or that we are dominated by the exozodiacal background. Thus, the intensity at the planet location consists of the planet PSF from Eq.~\ref{eq:shifted_planet} plus a uniform background, $I_b = I_{ss} + I_{sc} + I_{ez} + I_r$, where $I_{ss}$ is due to the residual suppressed starlight, $I_{sc}$ is due to scattered light, $I_{ez}$ is the exo-zodiacal dust background, and $I_r$ is the effective readout noise.  We assume that because of cosmic ray concerns the eventual signal will consist of a stack of many frames read out at a faster rate and thus it is sensible to treat the readout noise as absorbed into the uniform background.  We also assume that  the readout noise is small compared to the other sources of background noise.  In fact, it is expected that the dominant background source will be the exo-zodiacal light, perhaps as much as 3 to 10 times the brightness of the planet.


The photon arrival at each pixel is determined by a Poisson distribution.  That is, the probability of a given measurement, $z_{ij}(t)$ at pixel $(i,j)$ and time $t$ is given by:
\[
\mbox{Pr}[\mathbb{Z} = z_{ij}(t)] = e^{-I_{ij} t} \frac{(I_{ij} t)^{z_{ij}}}{z_{ij}!}
\]
where $I_{ij}$ is the mean arrival  rate of the photons, or intensity, at pixel $(i,j)$ in photons/sec.  Thus, if $z(t)$ is the actual measured number of counts at any pixel,
\begin{eqnarray}
\mathcal{E}\{z(t)\} &=& I t = \mu(t) \nonumber \\
\mathcal{E}\{(z(t)-\mu(t))^2\} &=&I t = \sigma^2(t) \nonumber
\end{eqnarray}
 %
where $I$ is the intensity at that pixel.

The photon arrival rate, or intensity, in photons/sec, at pixel $(i,j)$ is the integrated irradiance over the pixel area and a reference passband about a reference wavelength  $\lambda_r$, which for the planet is given by,  
\begin{equation}
I_{ij} =  \eta^2 I_p(\lambda_r) \int_{\lambda_r - \Delta\lambda/2}^{ \lambda_r + \Delta\lambda/2} {\int{\int_{\Delta \alpha_{ij}} {P(\lambda,u-\theta_x f,v - \theta_y f) d\lambda du dv}}}
\end{equation}
where $\Delta \alpha_{ij}$ represents the region covered by pixel $(i,j)$ and $\Delta \lambda$ is the reference wavelength band given by the largest band of interest.\footnote{For TPF-C this widest  band of interest, and the one used for this ideal reference case, is planned to be from 0.5 to 0.8 micron.  Strictly speaking, the variation of the field intensity, $I_p(\lambda)$, should be included in the integral over wavelength as the Planck curve has significant variations over this wide a band.  For simplicity, however, we are approximating the response curve as flat here for the purposes of developing the metric.  In the less idealized case where we measure over narrow bands this approximation becomes much better.}  In general there is not much more that can be done without a specific point spread function (and then these integrals can be difficult in closed form).  However, since we are seeking a simple and conservative metric, we can obtain a useful result by assuming  that the bandpass is small relative to $\lambda_r f/D$.  In this case, the pixel intensity reduces to the simpler form,
\begin{equation}
\label{eq:Iij}
I_{ij} \cong \eta^2 \Delta \lambda I_p(\lambda_r) \Delta \alpha \left [ \frac{1}{1 - (\Delta \lambda_r/2\lambda_r)^2} \right ] P_{ij}
\end{equation}
where the unitless quantity, $P_{ij} =  \frac{1}{\Delta \alpha} \int{\int_{\Delta \alpha_{ij}} {P(\lambda_r,u-\theta_x f,v - \theta_y f) du dv}}$, is the  normalized integral of the PSF over pixel $(i,j)$, and $\Delta \alpha$ is the pixel area.  Since we are considering small wavelength bands, we can safely drop the term in brackets in \refeqn{eq:Iij}. 

The final electron count rate at each pixel (also Poisson distributed), including background, is then,
\begin{equation}
\hat{I}_{ij} =  \epsilon \eta^2  \Delta \lambda I_p(\lambda_r) \Delta \alpha P(0,0)  \bar{P}_{ij}+ \epsilon \eta^2 \Delta \lambda I_b \Delta \alpha 
\end{equation}
where $\epsilon$ is the quantum efficiency of the camera and we have normalized the PSF by its value at (0,0) (so that $\bar{P}_{ij} = P_{ij}/P(0,0)$ and likewise $\bar{P}(\lambda,u,v) = P(\lambda,u,v)/P(\lambda,0,0)$).  We introduce the carat to differentiate between the photon and electron counts.

Finally, it will become useful later to utilize the contrast ratio, $Q$, first introduced by \cite{ref:Brown_Burrows}. The contrast ratio is defined as $I_p P(0,0)/I_b$.  In words, $Q$ is the ratio of the peak of the planetary image (in photon sec$^{-1}$ m$^{-2}$ $\mu$m$^{-1}$ ) to the local surface brightness of the background (in the same units).

It is often helpful, and very  common,  to non-dimensionalize the image plane coordinates.  Recalling that $P(\lambda_r,0,0) = A_c^2/\lambda_r^2 f^2$, the normalized PSF is given by,
\begin{equation}
\bar{P}(u,v) =\frac{P(u,v)}{P(0,0)} =  \frac{1}{A_c^2} \left | \int{\int_S{\mathcal{A}(x,y) e^{-j \frac{2\pi}{\lambda f}(xu + yv)} dx dy}} \right | ^2
\end{equation}
%

Next, introducing a reference length, $D$ (the side of a square aperture or the diamter of a circular aperture, for instance),  defining dimensionless pupil plane coordinates, $X=x/D, Y=y/D$, and a dimensionless angular position in the image plane in units of $\lambda/D$,
\[ \xi = \frac{Du}{\lambda f},\zeta = \frac{Dv}{\lambda f} \]
 gives the dimensionless form of the PSF,
 \begin{equation}
 \bar{P}(\xi, \zeta)  = \frac{D^4}{A_c^2} \left | \int{\int_{\bar{S}} {\mathcal{A}(X,Y) e^{-j 2\pi(X\xi + Y \zeta)} dX dY}} \right | ^2 
 \end{equation}
This equation can be modified slightly by introducing the definition $sD^2 \equiv A$.  That is, the entrance aperture area can be related to the square of the reference length by the constant $s$.  Thus, for example, for square apertures $s=1$, for circular apertures $s = \pi/4$, and for elliptical apertures $s=\pi r/4$, where $r$ is the aspect ratio of the ellipse.  The normalized PSF in this notation is given by,
 \begin{equation}
 \bar{P}(\xi, \zeta)  = \frac{A^2}{A_c^2} \frac{1}{s^2} \left | \int{\int_{\bar{S}} {\mathcal{A}(X,Y) e^{-j 2\pi(X\xi + Y \zeta)} dX dY}} \right | ^2 
 \end{equation}

Finally, $(u,v)$ is also replaced in the expression for $\bar{P}_{ij}$ to get the pixel integrated PSF in dimensionless units,
\begin{eqnarray}
\bar{P}_{ij} &=& \frac{\lambda_r^2 f^2 \Delta \bar{\alpha}}{\Delta \alpha D^2}\frac{1}{\Delta \bar{\alpha}} \int{\int_{\Delta \bar{\alpha}_{ij}} {\bar{P}(\xi,\zeta) d\xi d\zeta}}  \nonumber \\
&=& \frac{1}{\Delta \bar{\alpha}} \int{\int_{\Delta \bar{\alpha}_{ij}} {\bar{P}(\xi,\zeta) d\xi d\zeta}} 
\end{eqnarray}
where  $\Delta \bar{\alpha}$ is the dimensionless pixel area ($\Delta \bar{\alpha} = \frac{D^2}{\lambda_r^2 f^2}\Delta \alpha$).  Thus, as expected, $\bar{P}_{ij}$ can be computed in either dimensional or non-dimensional coordinates.  

Substituting for $P(0,0)$ and $\Delta \alpha$, the measurement equation can be re-written,
\begin{equation}
\label{eq:meas2}
\hat{I}_{ij}(t) =  \epsilon \eta^2  \Delta \lambda I_p(\lambda_r) \Delta \bar{\alpha} \frac{ A_c^2}{A}  s  \bar{P}_{ij}+ \epsilon \eta^2 \Delta \lambda I_b \Delta \alpha 
\end{equation}
Finally, we define the total throughput, $T=A_c/A$, the ratio of the exit pupil pseudo-area to entrance pupil area.  Substituting into Eq.~\ref{eq:meas2} leaves,
\begin{equation}
\hat{I}_{ij}(t) =  \epsilon \eta^2   \Delta \lambda I_p(\lambda_r) \Delta \bar{\alpha} T^2 A s  \bar{P}_{ij}+ \epsilon \eta^2 \Delta \lambda I_b \Delta \alpha 
\end{equation}

Thus, the intensity has been rewritten in terms of the total throughput and dimensionless pixel area. 

\section{Planet Detection}
\label{sec:detection}

In this section we examine  planet detection via a linear combination of pixel measurements.  The usual approach is to do PSF fitting where the irradiance of the planet is estimated by fitting a PSF to CCD image using standard photometric approaches \citep{Brown05}.  In other words, a planet PSF fit is attempted at every pixel and the results compared to find a statistically significant outlier.  A linear, unbiased procedure for PSF fitting is described in \cite{burrows03} and summarized in the appendix.  Here, we provide a probabilistic basis for using PSF fitting for planet detection and develop unambiguous metrics.  We also show that PSF fitting is, in fact, the optimal linear detection strategy.

\subsection{PSF Fitting}
\label{sec:psf_fitting}

As in \cite{burrows03}, we represent the Poisson counting process by a linear measurement model,
\begin{equation}
\label{eq:model}
z_{ij} = C_p \bar{P}_{ij} + C_b + \nu
\end{equation}
where $C_p = \epsilon\eta^2  \Delta \lambda I_p(\lambda_r) \Delta \bar{\alpha} T^2 sA t$,  $C_b = \epsilon \eta^2 \Delta \lambda I_b \Delta \alpha t$,  
$\nu$ represents the photon and readout noise, and $t$ is the integration time. 
Thus, $C_p$ is approximately the mean photon count at the pixel centered on the PSF (since $P_{00} \sim 1$) and $C_b$ is the mean background photon count at any single pixel.  

Given this model in Eq.~\ref{eq:model}, a variety of methods are available to estimate $C_p$, the planet irradiance.
The most common approach to photometry is via 
 PSF fitting (also called ``Matched Filtering'').  The standard, linear method  is described in \cite{burrows03}.  We summarize this derivation in the Appendix and compare to maximum likelihood approaches.    
 
 While \cite{burrows03} primarily  considers the background limited case, here we must assume that the planet signal and the background are of comparable intensity.  (It is possible that the exo-zodi could be as much as three to four times the intensity of the planet.  Nevertheless, this is not large enough to justify the background limited regime and thus it is best to include both the planet and background noises.  Note that \cite{Brown05} uses the background limited results of \cite{burrows03} yet includes the planet signal, resulting in an inconsistency in his results.)

While it is possible to consider nonlinear  maximum likelihood approaches to PSF fitting, these are complicated, provide minimal improvement (if any), and don't have convenient expressions for the estimate and variance.  We therefore continue to use the linear, unbiased estimator suggested by \cite{burrows03},
\begin{equation}
\label{eq:Cp_estimate}
\hat{C}_p = \frac{\sum_{ij} (z_{ij} - C_b)\bar{P}_{ij}}{\sum_{ij} \bar{P}^2_{ij}}
\end{equation}
where the sum is taken over some region $\Delta S$ of the image plane covered by the PSF. \footnote{Note that throughout we have assumed that the background is uniform in the summation region about the PSF (roughly 10 to 20 pixels).  We also assume the mean value is known in order to do the subtraction in Eq.~\ref{eq:Cp_estimate}.  The uniformity assumption is easily relaxed without changing the results by simply indexing $C_b$, though on this scale the exo-zodi should be very uniform.  However, if the background is unknown the problem becomes more complicated.  For telescope fixed background (speckle due to errors in the optics), the subtraction is accomplished by rotating the telescope and subtracting two measurements, at the expense of doubling the background noise.  If the zodi is uknown, then a simultaneous estimation is necessary; this complicates the results and will be treated in a future paper.}

For pixels with a planet, this is an unbiased estimate of the planet signal, $C_p$, with variance,
\begin{equation}
\label{eq:Cp_variance}
\sigma^2_{C_{p}} = \frac{C_p \sum_{ij} \bar{P}^3_{ij}}{(\sum_{ij} \bar{P}^2_{ij})^2} + \frac{C_b}{\sum_{ij} \bar{P}^2_{ij}}
\end{equation} 

For pixels without a planet, the expected value of   Eq.~\ref{eq:Cp_estimate} is zero and the variance is,
\begin{equation}
\label{eq:Cb_variance}
\sigma^2_{b} = \frac{C_b}{\sum_{ij} \bar{P}^2_{ij}}
\end{equation}

Planet detection is performed by calculating the estimate in Eq.~\ref{eq:Cp_estimate} at each pixel in the dark hole region.  For pixels without planets, the estimate will scatter about zero within the range defined by the variance in Eq.~\ref{eq:Cb_variance}.  For pixels with a planet, and a long enough integration time, the planet estimate should exceed this region about zero by some amount (see Fig.~\ref{fig:snr_sim} for an illustrative example).  We can plot the estimate for all pixels and look for where it exceeds the $1\sigma$ bounds defined by Eq.~\ref{eq:Cb_variance} by some factor, usually 5 or greater (a $5\sigma$ detection). 

\subsection{Detection Metric and Missed Detection Probability}
\label{sec:psffit_md}

 \cite{burrows03}, and later \cite{Brown05}, define the $SNR$ as the ratio of the actual planet irradiance, $C_p$ (the expected value of the estimate), and the standard deviation of the estimate error.  While this is a reasonable indicator of the quality of the estimate, it says little about how to make a detection decision.  The irradiance, $C_p$, is not the signal nor is the estimate variance the noise.  This approach to computing integration time also ignores that fact that the estimate of $C_p$ is also a random variable and thus has a specific, quantifiable probability of failing any detection test.  Finally, \cite{Brown05} uses both the planet and background photon noise in the definition of the $SNR$.  As just described, only the background is needed to perform a detection.

For the detection problem, we are interested in a metric that can be used to decide whether a planet exists in a given region.  This can be used to determine a maximum integration time before moving on.  The signal in this case is the estimate of the irradiance, $\hat{C}_p$ (from Eq.~\ref{eq:Cp_estimate}), and the noise is given by the background standard deviation on pixels without a planet (from Eq.~\ref{eq:Cb_variance}).  Thus, a detection occurs when the estimate exceeds the background by some amount.  The resulting signal-to-noise ratio metric, $\hat{C}_p/\sigma_b$, is a random variable and we can therefore define a confidence interval for the detection decision.  We choose an integration time such that the  probability that the $SNR$, assuming a planet exists, exceeds a given threshold, $K$, is equal to a desired value, P,
\begin{equation}
\label{eq:probability}
Pr \left \{ \frac{\hat{C}_p}{\sigma_b} > K | \mbox{a planet} \right \} = P
\end{equation}
For example, for 99\% confidence, $P=0.99$.  That is, if the correct integration time is used and no planet is seen, then we can say with 99\% confidence that there is no planet at that location (or at least, that there is no planet within $K \sigma_b$ of the background noise).

To compute this probability, we first  use Eqs.~\ref{eq:Cp_estimate} and \ref{eq:Cb_variance}, to find the $SNR$,
\begin{equation}
\frac{\hat{C}_p}{\sigma_b} = \frac{\sum_{ij} (z_{ij} - C_b)\bar{P}_{ij}}{\sqrt{C_b \sum_{ij} \bar{P}^2_{ij}}}
\label{eq:cp_sigma}
\end{equation}

Unfortunately, the probability mass function of $\hat{C}_p/\sigma_b$ is extremely difficult to compute using a Poisson process for the $z_{ij}$.  However, for the integration times being considered, the arrival rate is high enough that the Poisson process is well approximated by a Normal one and this assumption allows for a straightforward, and adequate, estimate of the integration time.  Also, by the Central Limit Theorem, the sum in Eq.~\ref{eq:cp_sigma} normalizes the  random variable.  From Eq.~\ref{eq:cp_sigma} the mean and variance are easily computed,
\begin{eqnarray}
\mu_{SNR} & = & \sqrt{C_p Q \sum \bar{P}^2_{ij}} \\
\sigma^2_{SNR} & = & 1+ \frac{Q \sum \bar{P}^3_{ij}}{ \sum \bar{P}^2_{ij}}
\end{eqnarray}
where we have used the definition of $Q$ introduced earlier.

This allows the probability in Eq.~\ref{eq:probability} to be rewritten in terms of a normalized variable, $Z$,
\begin{equation}
Pr\left \{ Z > \frac{K-\mu_{SNR}}{\sigma_{SNR}} | \mbox{a planet} \right \} = P
\end{equation}
where $Z$ is a standard normal random variable with zero mean and unit variance.  The detection criteria is then given by the probability integral,
\begin{eqnarray}
1 - \Phi \left ( \frac{K-\mu_{SNR}}{\sigma_{SNR}} \right ) &=& P \\
\mbox{or} \nonumber \\
\Phi \left ( \frac{K-\mu_{SNR}}{\sigma_{SNR}} \right ) &=& 1- P = P_{MD}
\end{eqnarray}
where $P_{MD} = 1 - P$ is the probability of a missed detection and $\Phi(z)$ is the gaussian distribution function,
\begin{equation}
\Phi(z) = Pr [Z \le z] = \int_{-\infty}^z \frac{1}{\sqrt{2\pi}} e^{\frac{-y^2}{2}} dy = \frac{1}{2} \left ( 1+\mbox{erf}(z/\sqrt{2}) \right )
\end{equation}

In other words, the integration time is chosen so that $\frac{K-\mu_{SNR}}{\sigma_{SNR}}$ reaches a value, $\gamma$, such that $\Phi(\gamma) = 1-P$.  For example, a 0.1\% missed detection rate ($P_{MD} = 0.001$) corresponds to $\gamma = -3.1$.  The detection criteria can now be written,
\begin{equation}
\frac{K - \sqrt{C_p Q \sum \bar{P}^2_{ij}}}{\sqrt{1+ \frac{Q \sum \bar{P}^3_{ij}}{ \sum \bar{P}^2_{ij}}}} = \gamma
\end{equation}
To solve for the needed integration time to achieve the desired false alarm and missed detection rates, this equation is first solved for $C_p$ (since $Q$ is independent of $t$) and then divided through to find the integration time,
\begin{equation}
t = \frac{1}{\epsilon \eta^2 \Delta \lambda I_p(\lambda_r) T^2 A s}  \frac{\left ( K - \gamma \sqrt{1+ \frac{Q \sum \bar{P}^3_{ij}}{ \sum \bar{P}^2_{ij}}} \right ) ^2}{Q \Delta \bar{\alpha} \sum \bar{P}_{ij}^2}
\label{eq:integration_time1}
\end{equation}
%

For consistency with \cite{ref:Brown_Burrows} and \cite{burrows03}, and for computational convenience, we make one more modification by introducing the {\em sharpness}.  Sharpness is defined in \cite{burrows03} as $\Psi = \sum_{ij} \bar{P}^2_{ij}/(\sum \bar{P}_{ij})^2$.  Using sharpness has the value of making the calculation less sensitive to the location of the PSF within a pixel.  Sharpness is a unique characteristic of the PSF of any coronagraph.  With this substitution, Eq.~\ref{eq:integration_time1} becomes,
\begin{equation}
\label{eq:integration_time}
t = \frac{1}{\beta} \frac{\left ( K - \gamma \sqrt{1 + \frac{\tilde{Q} \Xi}{\Psi}} \right )^2}{\tilde{Q} T_A   \Psi}
\end{equation}
where we have also defined $\beta = \epsilon \eta^2  \Delta \lambda I_p(\lambda_r) T A$, $\tilde{Q} \equiv Q \sum_{ij} \bar{P}_{ij}$,  $\Xi \equiv \frac{\sum \bar{P}_{ij}^3}{(\sum \bar{P}_{ij})^3}$ and $T_A \equiv T s  \Delta \bar{\alpha} \sum \bar{P}_{ij} = T s \int \int _{\Delta S} \bar{P}(\xi,\zeta) d\xi d\zeta$.  
$\tilde{Q}$ is the ratio of the total photon count from the planet over the integration area, $\Delta S$, to the number of background counts in a single pixel.  
$T_A$ is what we called the 
{\em Airy Throughput} in \cite{VSK03}.  It is the throughput times the integral of the normalized PSF over the integration area $\Delta S$. It is straightforward to show, using Parseval's Theorem, that $s \int \int _{\Delta S} \bar{P}(\xi,\zeta) d\xi d\zeta$  is one for any open aperture assuming that $\Delta S$ covers the entire image plane.  For the expected smaller integration area, $T_A$ is always less than $T$.  It is thus the fraction of the total throughput achieved for the given region of the PSF used in detection.  This difference can be significant.  For an open, circular aperture, the Airy throughput corresponding to the first null is only 84.5\%.  If only the FWHM is used, then this drops to 53\%, almost a factor of 2 in integration time.  For shaped pupils, this is also important since most shaped pupil PSFs move a significant portion of the light outside the {\em outer working angle}; this is light not used in the planet detection.

\begin{figure}
\plottwo{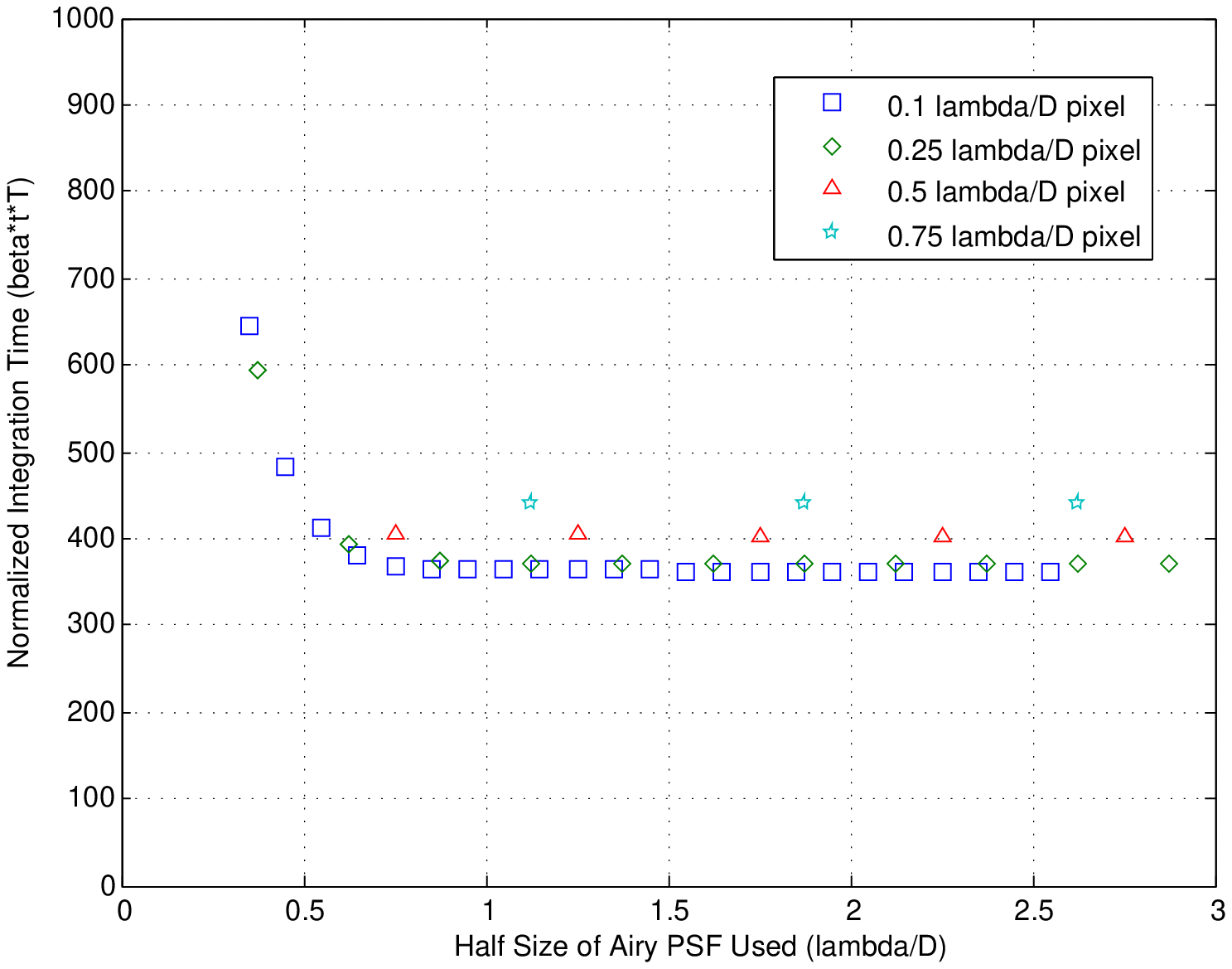}{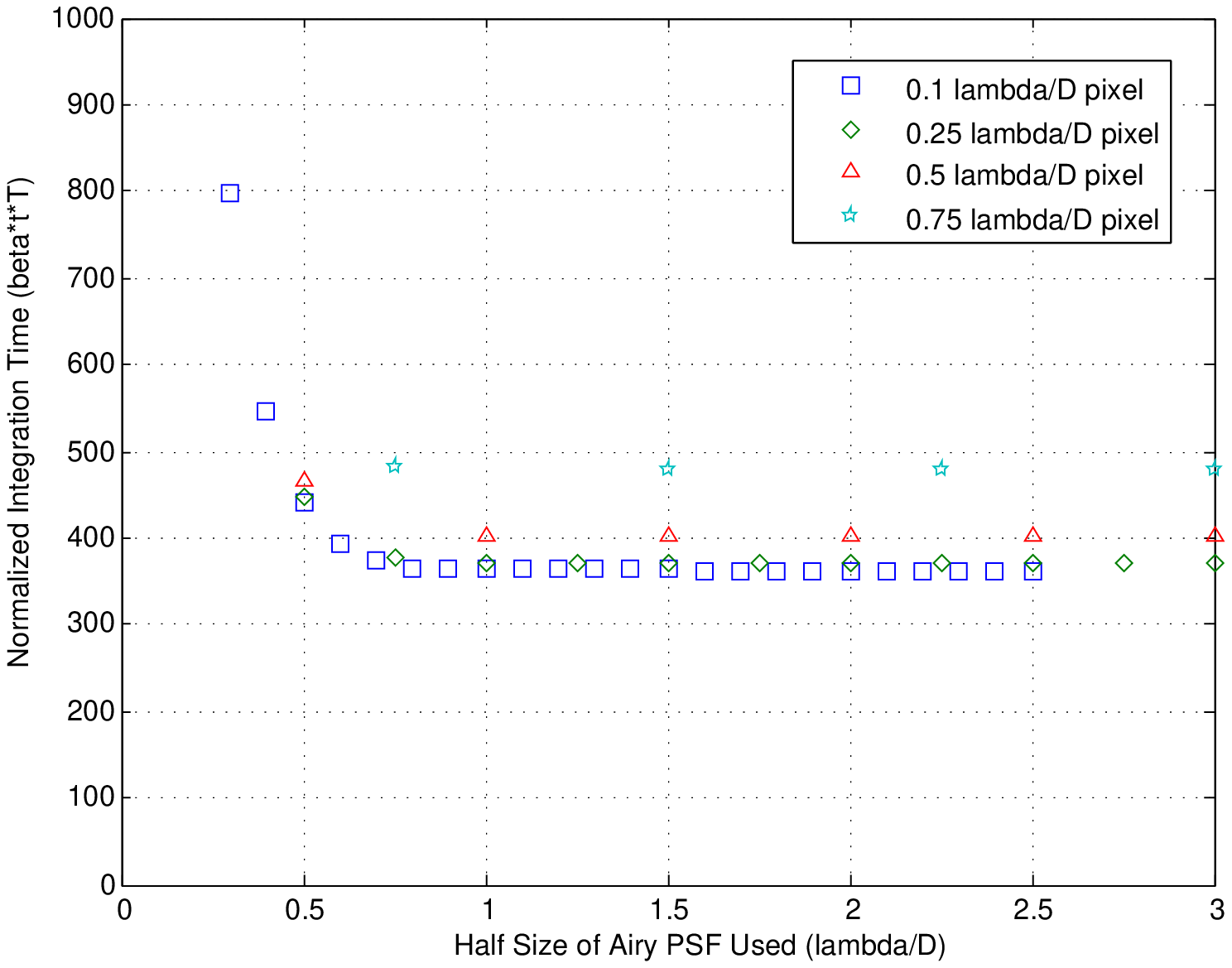}
\caption{Normalized integration time versus size of Airy PSF used in detection for various pixel sizes.  {\em Left} PSF centered on single pixel.  {\em Right} PSF centered on corner of four pixels.}
\label{fig:pixelsizes}
\end{figure}

This trend of the Airy Throughput would imply that it is always best to use as many pixels as possible to cover as much of the planet's PSF as possible.  However, the remaining terms that depend upon $\tilde{Q}$, $\Psi$, and $\Xi$ have the opposite trend, suggesting that a smaller number of pixels around the central core is better (since more pixels tend to include more background light but less planet light because of the drop off of the PSF).  The result is that beyond a certain region of the core there is no improvement in integration time.  This region size depends upon the specific PSF and the size of the pixels.  For an Airy PSF it is roughly 0.75  $\lambda/D$.  Figure~\ref{fig:pixelsizes} shows two plots of the normalized integration time ($\beta t T$) versus width of the core used in the detection for four different pixel sizes and a background limited detection, where $Q = 1/3$.  While there is an advantage to oversampling the PSF, the gain is slight.  For smaller background, and larger $Q$, slightly less of the core is necessary but the difference in integration time is also small.

This also raises the question of the location of the PSF within the pixel.  Up to now, this has been ignored.  Figure~\ref{fig:pixelsizes} shows that by  using sharpness the integration time expression is reasonably insensitive to the PSF location relative to the pixel center.  However, in practice, the estimate of $C_p$ in Eq.~\ref{eq:Cp_estimate} assumes that $P_{ij}$ is known at every pixel, introducing an ambiguity.  One solution is to attempt a fit for many values of $P_{ij}$ as the PSF moves relative to a pixel center.  Alternatively, we could compute the sensitivity of the detection to errors in $P_{ij}$ and increase the integration time accordingly.  Finally, we could treat the actual value of $P_{ij}$ as a random variable with uniform distribution between its value for the PSF landing on a pixel center and its value for the PSF landing on a pixel edge.  This probability distribution can then be included in the probability calculation in Eq.~\ref{eq:probability}.   We defer this approach to future work.

\subsection{Choice of K and False Alarm Probability}
\label{sec:psffit_fa}

Equation~\ref{eq:integration_time} provides the integration time to achieve a desired small missed detection rate for a given detection threshold, $K$.  However, the choice of $K$ still seems arbitrary.  Most astronomical applications select $K$ between 5 and 10 (while ignoring the missed detection probability embedded in $\gamma$).  In reality, $K$ has a probabilistic basis as well; it is a measure of the probability of false alarms.  In fact, the best approach is to pick the integration time, $t$, and the threshold, $K$, to achieve a desired balance between the two types of errors.  

The probability of false alarms is given by an expression similar to that in Eq.~\ref{eq:probability},
\begin{equation}
Pr \left \{ \frac{\hat{C}_p}{\sigma_b} > K | \mbox{no planet} \right \} = P_{FA}
\label{eq:p_fa}
\end{equation}
The expression for $\hat{C}_p/\sigma_b$ is the same as that in Eq.~\ref{eq:cp_sigma}, but since there is no planet the mean and variance differ,
\begin{eqnarray}
\mu_{FA} & = & 0\\
\sigma^2_{FA} & = & 1 
\end{eqnarray}
The $SNR$ when there is no planet in a pixel is, in fact, well approximated by a standard normal random variable.  The false alarm probability is then given simply by,
\begin{equation}
\Phi(K) = 1 - P_{FA}
\label{eq:phi_fa}
\end{equation}
Thus, Eq.~\ref{eq:phi_fa} is used to select the threshold, $K$, corresponding to a desired probability of false alarms and Eq.~\ref{eq:integration_time} is then used to select the integration time to achieve a desired probability of missed detections.

Since we expect at best to only find one planet per system surveyed, most pixels will not have a planet.  This places a high premium on keeping the false alarm rate low (a large $K$), otherwise much  time would be wasted on follow up of false planets.  Nevertheless, a $K$ greater than 5 results in a false  alarm probability lower than necessary (e.g., $\Phi(5) = 0.9999997$, a false alarm probability of 3e-7).  The result is a potentially longer integration time than necessary, dramatically affecting the mission.  A reasonable approach is to select the false alarm probability such that there is less than one false alarm per observed system.   For instance, for a TPF mission that surveys 100 systems, a total false alarm rate of 1\% would result in 1 false alarm over the course of the mission.  Since each planet detection will be followed by deep surveys for characterization, the false alarm would be quickly discovered and have minimal impact on mission life.    If we assume on the order of 3000 measurements in a given system, we therefore require that $P_{FA} \times 3000 \sim 0.1$ or $P_{FA} \sim 3 \times 10^{-5}$.  This corresponds to $K = 4$.

The missed detection probability is critical for a mission with so few surveyed systems and unknown planet occurrence rate, $\eta_\oplus$.  It  should be made low enough to avoid a single missed detection over the course of the mission.  A missed detection probability of less than 0.1\% is most likely needed, requiring $\gamma = -3.1$.

With these numbers, the integration time equation becomes,
\begin{equation}
\label{eq:PSF_fitting_integration_time}
t = \frac{1}{\beta} \frac{\left ( 4 +3.1 \sqrt{1 + \frac{\tilde{Q} \Xi}{\Psi}} \right )^2}{\tilde{Q}  T_A \Psi}
\end{equation}

There are a number of notable differences between this expression for integration time and that in \cite{Brown05}.
First, \cite{Brown05} selected the $SNR$ threshold by implicitly only considering the false alarm probability.  He ignored the missed detection probability ($\gamma$) and the potentially significant contribution to the integration time in Eq.~\ref{eq:integration_time}. (Physically, this corresponds to basing the detection decision only on the mean value of the $SNR$ which leaves a significant probability that a particular $SNR$ may fall below the threshold.  Mathematically, this means $\gamma=0$ which corresponds to a missed detection probability of 50\%.)  \cite{Brown05} also neglected to account for the possible reduction in Airy throughput due to only using the center of the PSF.

Another difference is that \cite{Brown05} included the planet photon noise in the $SNR$ expression.  We describe above why only the background noise should be used in the $SNR$ definition.   

Also note that the needed integration time depends upon the actual planet irradiance (in $\beta$).  Strictly speaking, this is itself a random variable as the planet irradiance could vary quite widely among systems.  While it is possible to include a density function for $I_p$ and marginalize the $SNR$, this makes the problem substantially more complicated.  It is also not clear that an {\em a priori} density for the planet irradiance can easily be found.  A more reasonable approach is to use a worst case irradiance in Eq.~\ref{eq:integration_time} for the most conservative integration time.  This will become important again in the Bayesian analysis later.
 
 \subsection{Numerical Example}

In this section we present a simple numerical calculation of the integration time for a typical system.  For simplicity, we ignore the ellipticity of the TPF-C entrance aperture.   To begin, we also ignore the coronagraph (and any affect on the PSF)  and consider the shortest possible integration time to detect a planet given the telescope aperture.  In other words, $A_c = A$ , the entrance aperture area.  I assume the entrance area, $A$, is the same as the 8m x 3.5m TPF-C design (22 m$^2$).  
Assuming a 5th magnitude star and a planet with $\Delta mag$ of 25, the irradiance is approximately $I_p = 9.5 \times 10^{-9} \mbox{ photons cm}^{-2} \mbox{ nm}^{-1} \mbox{ sec}^{-1}$.  With a QE of 0.8 $\Delta \lambda = 100$ nm, and $\eta^2 = 0.33$, $\beta = 0.06$ photons/sec.  The integration time then becomes,
\begin{equation}
t \cong 18 \frac{\left ( 4 +3.1 \sqrt{1 + \frac{\tilde{Q} \Xi}{\Psi}} \right )^2}{\tilde{Q}  T_A \Psi}
\end{equation}
Assuming a fairly conservative value of $Q=1/3$ (large exo-zodi), we can use Fig.~\ref{fig:pixelsizes} 
to obtain the normalized integration of roughly 400 for a critically sampled Airy function.  The resulting integration time becomes 7200 seconds or 2 hours.

A more realistic case would include the effect of a Lyot stop (or shaped pupil).  The Lyot stop reduces $T$ and $T_A$ (by 0.3 for the current design).    If we assume that a detector with larger pixels is used so that the Airy function is still critically sampled, the  sharpness, $\Psi$, $\Xi$, and $\tilde{Q}$ are all unchanged.   The result is an increase in the integration time by a factor of $1/0.3^2$ or about 11.  This implies a required integration time of 22 hours.

\subsection{General Linear Detection \& Optimality}


The specific irradiance estimate in Eq.~\ref{eq:Cp_estimate} can be easily generalized into a form that covers all possible linear processing techniques,
\begin{equation}
Z = \sum_{ij} W_{ij} (z_{ij} -C_b)
\end{equation}
 for some unknown set of weights, $W_{ij}$, over a region $\Delta S$ of the image.  Following the same process as before, the signal-to-noise ratio is,
 \begin{equation}
 \frac{Z}{\sigma_b} = \frac{\sum_{ij} W_{ij} (z_{ij} - C_b)}{\sqrt{C_b \sum_{ij} W^2_{ij}}}
 \label{eq:test_statistic}
 \end{equation}
 with mean $\mu_{SNR} = \sqrt{C_p Q}\sum_{ij} W_{ij} \bar{P}_{ij}/ \sqrt{\sum_{ij} W^2_{ij}}$ and variance $\sigma^2_{SNR} = 1 + \frac{Q \sum W^2_{ij} \bar {P}_{ij}}{\sum W^2_{ij}}$.  By the same probabilistic argument, the detection condition becomes,
 \begin{equation}
 \frac{K-\frac{\sqrt{C_p Q}\sum W_{ij} \bar {P}_{ij}}{\sqrt{\sum W^2_{ij}}}}{\sqrt{1 + \frac{Q \sum W^2_{ij} \bar {P}_{ij}}{\sum W^2_{ij}}}} = \gamma
 \end{equation}
 As before, this can be solved for $C_P$ and then $t$,
 \begin{equation}
 t = \frac{1}{\beta} \frac{\left ( K - \gamma \sqrt{1 + \frac{Q \sum W^2_{ij} \bar {P}_{ij}}{\sum W^2_{ij}}} \right ) ^2 \sum W_{ij}^2 }{Q \Delta \bar{\alpha} (\sum W_{ij} \bar{P}_{ij} )^2}
 \label{eq:general}
 \end{equation}

 One observation from Eq.~\ref{eq:general} is that the weights can be multiplied by an arbitrary constant without effecting the integration time (and thus the detection).  This is because any constant weight divides out when forming the $SNR$ statistic.  Thus, while the constant term $\sum_{ij} \bar{P}^2_{ij}$ in Eq.~\ref{eq:Cp_estimate} is important for an unbiased estimate of $C_p$ (photometry), it is not necessary for a simple detection. 
 
 %

Equation~\ref{eq:general} can be used to find the optimal weights resulting in a minimum integration time.  These turn out to be very close to PSF fitting in the last section.  

This minimum is found by differentiating Eq.~\ref{eq:general} with respect to $W_{ij}$ and setting equal to zero.  Unfortunately, the exact problem is difficult to solve because of the term in the radical.  However, for $Q \ll 1$, this term is small and the integration time is very well approximated by, 
\begin{equation}
\label{eq:general_integration_time}
 t \cong \frac{1}{\beta }  \frac{(K-\gamma)^2 \sum W^2_{ij}}{Q \Delta \bar{\alpha} (\sum W_{ij}\bar{P}_{ij})^2}
\end{equation}
The resulting minimum is found from,
\begin{equation}
\frac{\partial}{\partial W_{ij}}\left ( \frac{\sum W^2_{ij}}{(\sum W_{ij}\bar{P}_{ij})^2} \right ) = 0
\end{equation}

Completing the differentiation results in the condition for $W_{ij}$,
\begin{equation}
W_{ij} = P_{ij} \frac{\sum W^2_{ij}}{\sum W_{ij} \bar{P}_{ij}}
\end{equation}

While difficult to solve in general, it is easy to show that this equation is satisfied by the PSF fitting weights used in the previous section, $W_{ij} = \bar{P}_{ij}$.   Thus, for small $Q$ (large background), the PSF fitting approach is the approximately optimal linear detection technique.  Any improvement from searching for the exact optimal weights is only very slight.

\section{Bayesian Detection}

We showed above that PSF fitting for detection was the optimal, that is, shortest time, detection among all linear algorithms.  It can also be shown that under certain assumptions it achieves the Cramer-Rao bound and is thus the best possible estimate of the irraadiance.  However, for the detection problem we are not necessarily interested in planetary photometry, just a determination of whether a planet exists.  PSF Fitting may then provide more information than is necessary.  It is reasonable, therefore to question whether there might be faster algorithms for making the detection decision.   In this section we explore an alternative  hypothesis testing approach using Bayesian methods using the same measurement model as in Section 3 in an effort to answer this question.

\subsection{Detection Criteron}

To determine whether a potential planet
is present in a set of pixels $\Delta S$, centered at pixel $n$, we compare the probabilities of
the following two alternate hypotheses:
\begin{eqnarray*}
H_{0}^n &:&\mbox{there is no planet in the }n-th\mbox{ pixel,} \\
H_{1}^n &:&\mbox{there is a planet in the }n-th\mbox{ pixel.}
\end{eqnarray*}
Each hypothesis, $H_i$, has the associated  model $M_i$:
\begin{eqnarray*}
M_{0}^n &:& \lambda_{ij} =  C_b \\
M_{1}^n &:& \lambda_{ij} = C_p\bar{P}_{ij} + C_b
\end{eqnarray*}
where $\lambda_{ij}$ is the mean arrival rate of photons in the  Poisson distribution at pixel $(i,j)$ located in $\Delta S$.  

We now compare the posterior probabilities
of the two models conditioned on the collected data and
decide that there is a planet in pixel $n$ if the \emph{odds
ratio}, $O_{10}^n$, favors $H_{1}^n$ over $H_{0}^n$:
\begin{equation}
O_{10}^n =\frac{p\left( M_{1}^n|Z\right) }{p(M_{0}^n|Z)}=\frac{
p\left( M_{1}^n\right) p\left( Z|M_{1}^n\right)
}{p(M_{0}^n)p\left( Z|M_{0}^n\right) } = R \frac{p\left( Z|M_{1}^n\right)}{p\left( Z|M_{0}^n\right) }
\label{defO10}
\end{equation}
where $Z$ is the set of collected measurements and we have used Bayes' rule to write the odds ratio as the
product of the ratio of prior probabilities,  $R = p\left( M_{1}^n\right)/
p(M_{0}^n)$, and the ratio of likelihoods, $p\left( Z|M_{1}^n\right)/
p\left( Z|M_{0}^n\right)$.  The dependence on the ratio of prior probabilities of there being a planet is troubling at first since it appears that the effectiveness of the test can be greatly affected by this essentially unknown quantity (alternatively, one can view this as perhaps an attractive feature as prior information on the frequency becomes available).  We will show later that this dependence is easily removed by balancing the sensitivity between missed detections and false alarms.

One possible and simple decision criterion would be that $O_{10}^n>1$
favors $H_1^n$, thus indicating that  there is a planet whose true location
$\zeta^*$ is inside $\Delta S$.  However, a better approach to the decision is to base it  on a different value of the ratio derived from particular weighting criteria.  This is called a ``Loss-based decision''.  The specific value  of the Odds Ratio for the decision is selected to balance the relative priorities between missed detections and false alarms (note the similarity to the PSF fitting test above).  In most applications this process is used to optimize the power of the test (via the Receiver Operator Curve or ROC).  Here, however, as  discussed above, there is a large incentive to reduce the probability of false alarms well below that of missed detections rather than balance the two types of errors.  In what follows, analytical expressions are developed to quantitatively trade off these two probabilities.

\subsection{Likelihoods and Priors}


In general, the probability of a given data set depends on various physical parameters of the system. For instance, the mean photon rate in the Poisson distribution depends upon the irradiance of the planet and its location in the pixel.  As mentioned in the previous section, this is, strictly speaking, a random variable.  The usual approach to computing the likelihood functions is to {\em marginalize} the probabilities to remove the parameter dependence altogether.  However, given our lack of knowledge regarding the planet distribution, and that our goal herein is to simply find a conservative estimate of the needed integration time for detection, it is more sensible to begin with a worst case analysis.  Thus, we assume the parameters (such as planet irradiance) are known and take on their most pessimistic value (for instance, for planets below a certain limiting $\Delta mag$ we assume a detection is impossible).  If an actual planet turns out to be brighter, then we have simply integrated longer than necessary, but we will not miss the detection.  In short, we trade conservatism for our lack of knowledge of the specific distribution of the priors.

The odds ratio now becomes the ratio of the likelihood functions times the prior probabilities,
\begin{equation}
O_{10}^n = R \frac{L_1}{L_0}
\label{eq:odds_ratio}
\end{equation}
where the likelihood functions are given by $L_1 = p(Z|M_1^n)$ and $L_0 = p(Z|M_0^n)$.   Each of these probability distributions per pixel are Poisson with parameter $\lambda_{ij}$,
\begin{equation}
L_l = \prod_{ij} \frac{1}{z_{ij}!} \lambda_{ij}^{z_{ij}} e^{-\lambda_{ij}}
\end{equation}
Thus, for a set of data per pixel, $z_{ij}$, we form the Odds Ratio using our worst  case value for the planet irradiance, $\bar{C}_p$, and the contrast ratio, $\bar{Q}$, where I have used the bar to contrast the value used in the detection algorithm from the true values, $C_p$ and $Q$.  The background irradiance, $C_b$, is still assumed to be known (and uniform).  Substituting into Eq.~\ref{eq:odds_ratio}, the Odds ratio becomes,
\begin{equation}
O_{10}^n = Re^{-\bar{C}_p \Delta P} \prod_{ij} (1 + \bar{Q} \bar{P}_{ij})^{z_{ij}}
\end{equation}
where $\Delta  P = \sum_{ij} \bar{P}_{ij}$.  It is often more convenient to use the log of the Odds Ratio as the test statistic,
\begin{equation}
\log O_{10}^n = \log R -\bar{C}_p \Delta P + \sum_{ij} z_{ij} \log(1+\bar{Q}\bar{P}_{ij})
\label{eq:log_odds_ratio}
\end{equation}

The two constant terms don't affect the statistical test, so it is easier to define a new test statistic that removes them,
\begin{equation}
\chi^n =  \sum_{ij} z_{ij} \log(1+\bar{Q}\bar{P}_{ij})
\end{equation}
Unlike in PSF fitting, not only do we need to know $P_{ij}$, we also need to know $Q$.  This begs the question of the sensitivity of the method to errors in our assumed knowledge, $\bar{Q}$.  As mentioned above, for now we use a worst case approach, using the smallest value of $I_p$ expected (recall that in both of these approaches the background is assumed known).  We explore the validity of this approach later.

The test statistic, $\chi$, depends upon the measured data and assumed, worst case parameters.  What remains is to find a threshold value for $\chi$ to decide between the  two hypotheses.

\subsection{Hypothesis Tests}

Again, the Loss based decision is made based upon the value of the test statistic.  That is, if $\chi^n > \gamma$ then we reject $H_0$ and decide that a planet is present in $\Delta S$.  Otherwise, we accept $H_0$ and assume no planet.   $\gamma $ is selected to balance the two types of errors.

\subsubsection{Missed Detections}
The probability of a missed detection can be written formally as,
\begin{eqnarray}
Pr\{ \chi^n < \gamma | M_1^n\} & = & P_{MD} \nonumber \\
\mbox{or} \nonumber \\
Pr\{\sum z_{ij} \log(1 + \bar{Q}\bar{P}_{ij}) < \gamma  | M_1^n\} & = & P_{MD}
\label{eq:md}
\end{eqnarray}

To compare to PSF fitting, we look for an expression that allows us to pick an integration time to achieve a certain desired probability of missed detection, $P_{MD}$.  Unfortunately, even knowing that $z_k$ is Poisson, forming the probability distribution in Eq.~\ref{eq:md} is formidable.  For a precise result, we must resort to Monte Carlo simulations.  However, for the expected integration times and arrival rates here, the Poisson distribution is close to Gaussian and, as in Section 4, this assumption is useful for obtaining integration time estimates.  Thus, letting $B_{ij} = \log(1+\bar{Q}\bar{P}_{ij})$, the probability in Eq.~\ref{eq:md} can be rewritten in terms of a standard normal  random variable,
\begin{equation}
Pr\{ Z < \frac{\gamma  - \mu_Z}{\sigma_Z} \} = P_{MD}
\end{equation}
where $\mu_Z = C_p \sum B_{ij} (\bar{P}_{ij} + 1/Q)$ and $\sigma^2_Z = C_p \sum B_{ij}^2 (\bar{P}_{ij} + 1/Q)$ and Z is normally distributed with mean 0 and standard deviation 1.  Thus, the missed detection equation becomes,
\begin{equation}
\Phi\left ( \frac{\gamma  - \mu_Z}{\sigma_Z} \right ) = P_{MD}
\end{equation}
where, as with PSF fitting, the desired missed detection probability, $P_{MD}$, is picked {\em a priori}.  Letting $\alpha$ be the value of the argument such that $\Phi(\alpha) = P_{MD}$, the integration time criteria becomes,
\begin{equation}
\label{eq:bayesian_time}
 \frac{\gamma  - \mu_Z}{\sigma_Z} = \alpha
 \end{equation}
 In other words, for a desired missed detection probability, the integration time is selected such that Eq.~\ref{eq:bayesian_time} is satisfied. 
 
 Substituting for $\mu_Z$ and $\sigma_Z$ leaves,
 \begin{equation}
 \gamma  - C_p \sum B_{ij} (\bar{P}_{ij} + 1/Q) = \alpha \sqrt{C_p \sum B_{ij}^2 (\bar{P}_{ij} + 1/Q)}
 \label{eq:missed_detection}
 \end{equation}
This equation has two unknowns, however, $\gamma$ and $t$.  As before, the false alarm rate provides a criteria for the threshold, $\gamma$, that can be substituted back in to find the integration time.
 
 \subsubsection{False Alarms}
 The probability of a false alarm is similar to Eq.~\ref{eq:md},
 \begin{eqnarray}
 Pr\{ \chi^n > \gamma | M_0^n\} & = & P_{FA} \nonumber \\
\mbox{or} \nonumber \\
Pr\{\sum z_{ij} \log(1 + \bar{Q}\bar{P}_{ij}) > \gamma  | M_0^n\} & = & P_{FA}
\label{eq:fa}
\end{eqnarray}
As before, this is rewritten in terms of a standard normal random variable,
\begin{equation}
Pr\{ Z' > \frac{\gamma  - \mu_{Z'}}{\sigma_{Z'}} \} = 1-\Phi \left(  \frac{\gamma  - \mu_{Z'}}{\sigma_{Z'}} \right ) = P_{FA}
\end{equation}
where here $\mu_{Z'} = \sum B_{ij} C_b$ and $\sigma^2_{Z'} = \sum B^2_{ij} C_b$.  Letting $\delta$ be the value of the argument such that $\Phi(\delta) = 1- P_{FA}$ leaves,
\begin{equation}
\frac{\gamma  - \mu_{Z'}}{\sigma_{Z'}} = \delta
\end{equation}
or,
\begin{equation}
\gamma = \delta \sqrt{\sum B_{ij}^2 C_b} + \sum B_{ij} C_b 
\label{eq:gamma}
\end{equation}
This equation provides the detection threshold test for $\chi^n$ for a given false alarm probability (embodied in the value for $\delta$).
 
 \subsubsection{Integration Time}
 
 The Bayesian integration time can now be found by using the false alarm based expression for $\gamma$ in Eq.~\ref{eq:gamma} in Eq.~\ref{eq:missed_detection},
 \begin{equation}
 C_p \left [ \frac{\sum B_{ij}}{Q} - \sum B_{ij}(\bar{P}_{ij} + 1/Q) \right ] = \alpha \sqrt{C_p \sum B^2_{ij}(\bar{P}_{ij} + 1/Q)} - \delta \sqrt{C_p \frac{\sum B^2_{ij}}{Q}}
 \end{equation}
 This can be used to solve for $C_p$ as before and subsequently the Bayesian integration time,
 \begin{equation}
 t_B =  \frac{1}{\epsilon  \Delta \lambda I_p(\lambda_r) T A_c s}  \frac{ \left (  \alpha \sqrt{\sum B^2_{ij}(Q\bar{P}_{ij} + 1)} - \delta \sqrt{\sum B^2_{ij}} \right ) ^2}{Q \Delta \bar{\alpha} \left (\sum B_{ij} \bar{P}_{ij} \right )^2}
 \end{equation}

 Finally, we can use the same renormalization as before and write the Bayesian integration time in terms of $\beta$, $T_A$, and $\tilde{Q}$,
 \begin{equation}
 \label{eq:bayesian_integration_time}
 t_B = \frac{1}{\beta}  \frac{ \left (  \alpha \sqrt{\sum B^2_{ij}(\tilde{Q}\bar{P}_{ij}/\sum \bar{P}_{ij} + 1)} - \delta \sqrt{\sum B^2_{ij}} \right ) ^2}{\tilde{Q} T_A \left (\sum B_{ij} \bar{P}_{ij}/\sum \bar{P}_{ij} \right )^2}
\end{equation}
 
 Surprisingly, for background limited systems where $Q$ is small, there is no improvement in integration time with the Bayesian approach.   This is rather easy to show using Eq.~\ref{eq:bayesian_integration_time}.  For small $Q$, $B_{ij} = \log(1+\bar{Q}P_{ij}) \cong \bar{Q}P_{ij}$.  With this subsitution, Eq.~\ref{eq:bayesian_integration_time} becomes exactly Eq.~\ref{eq:integration_time}. 
 Thus, this Bayesian approach, where we assumed we knew the background and used a worst case value of $I_p$ (i.e., $Q$), provides no integration time benefit over PSF fitting for small $Q$ (large background).  Additionally, one would expect a poorer performance (that is, a slight increase in the probability of false alarms and missed detections) when $\bar{Q} \neq Q$.  Also, this approach suffers from the same ambiguity regarding the location of the PSF relative to a pixel center as PSF fitting, as $P_{ij}$ has been assumed known throughout.
 
 One of the attractive features of the Bayesian approach is that it provides a methodology for handling these unknowns.  For example, the uncertainty in $\bar{Q}$ can be removed by assuming a distribution for $I_p$ and marginalizing the likelihood function.  Likewise, the ambiguity with regard to pixel size can also be removed via marginalization.  Finally, we can study the case where the background is not known by again marginalizing over a distribution for the background photon count.  An unknown background can also be treated in PSF fitting by simultaneously estimating the planet irradiance and background, at the cost of much  greater complexity and much longer integration times.  We defer these comparisons to a future paper.
  
 \section{Monte Carlo Simulation}
 
\begin{figure}
 \plottwo{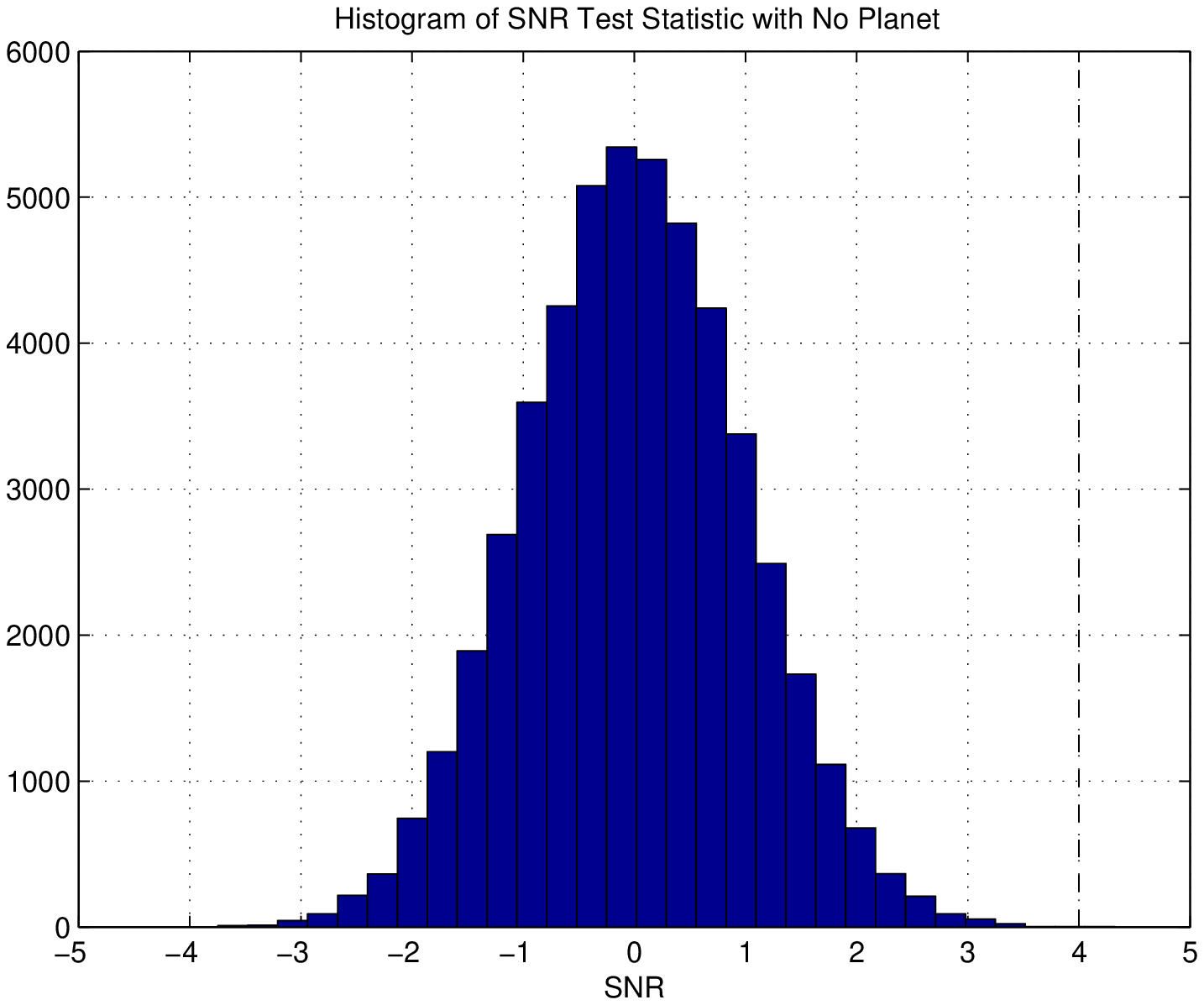}{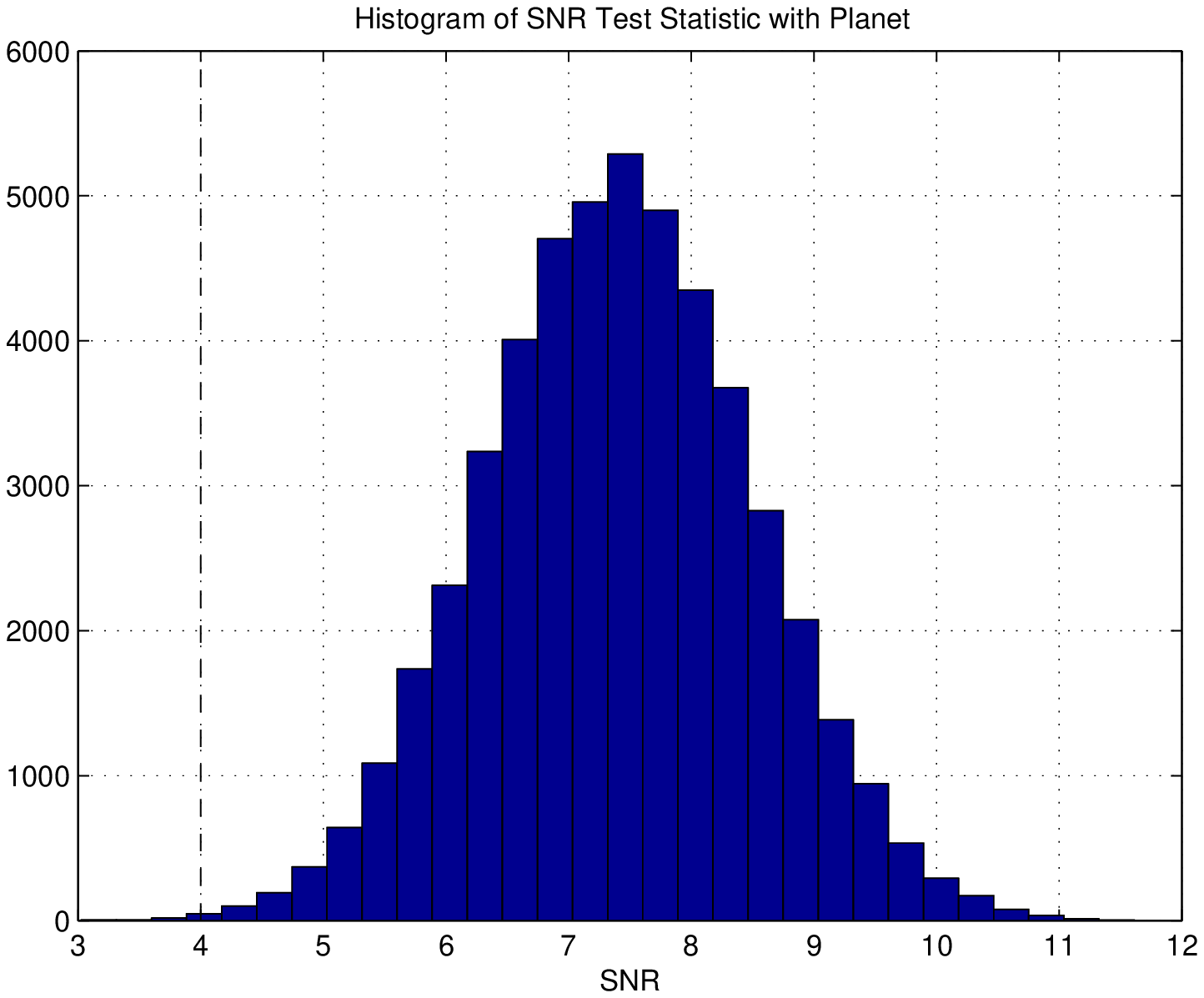}
 \caption{Histograms of 50,000 Poisson simulations of the PSF Fitting signal-to-noise ratio test statistic for each case, pixels without a planet ({\em Left}) and pixels with a planet ({\em Right}).  Dotted line indicates the threshold test value, $K$.}
 \label{fig:monte_snr}
 \end{figure}

\begin{figure}
 \plottwo{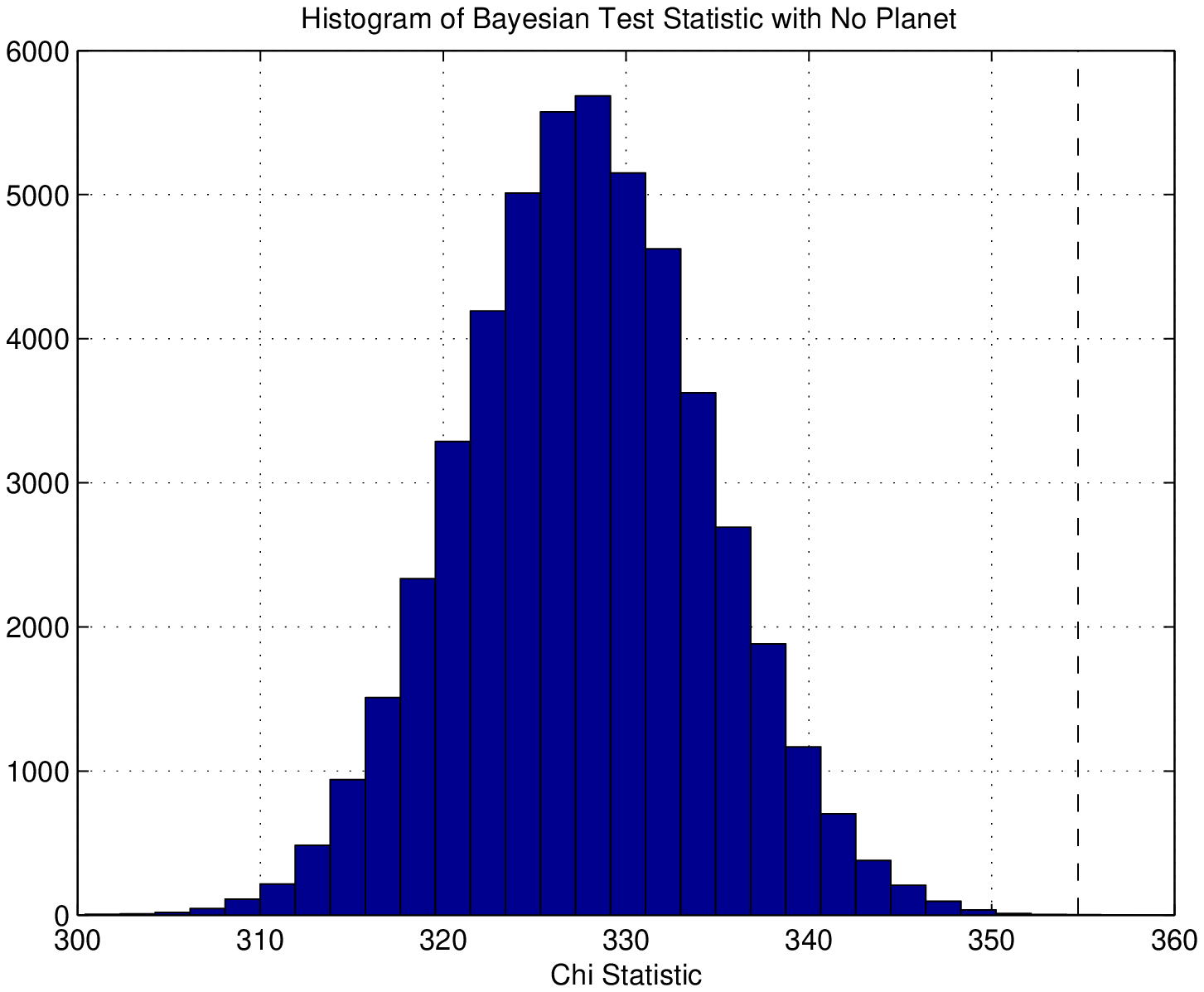}{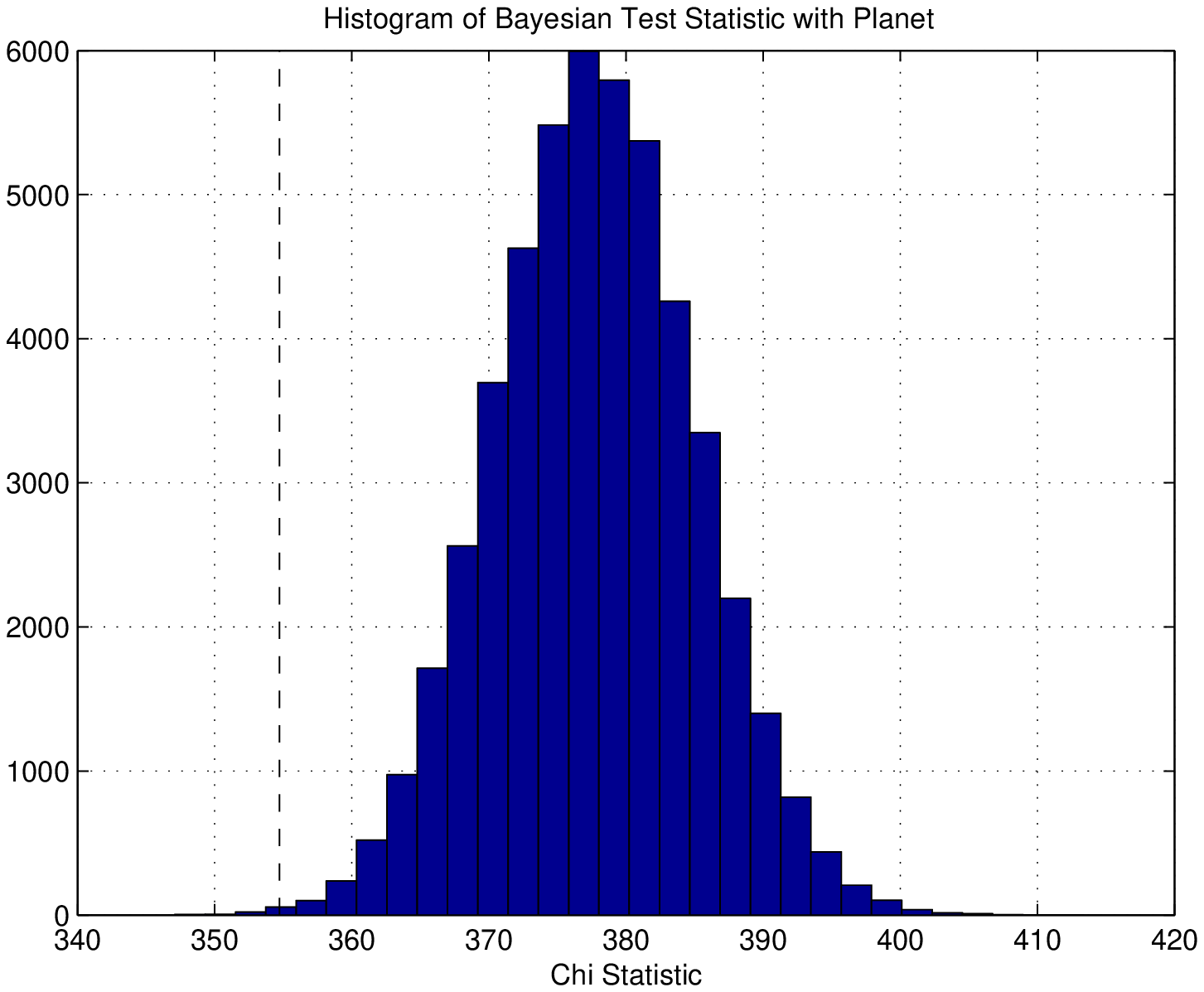}
 \caption{Histograms of 50,000 Poisson simulations of the Bayesian test statistic for each case, pixels without a planet ({\em Left}) and pixels with a planet ({\em Right}). Dotted line indicates the threshold test value, $\gamma$.}
  \label{fig:monte_chi}
 \end{figure}

 As a straightforward verification, we ran 50,000 monte carlo runs for each case, with a planet and without.  The Poisson arrival rate was computed using the integration time values computed in Eqs.~\ref{eq:PSF_fitting_integration_time} and \ref{eq:bayesian_integration_time}.  An Airy PSF was used for all simulations and $Q$ was taken as 1/3.   Figure~\ref{fig:monte_snr} shows histograms of the results of the signal-to-noise test statistic using PSF fitting and Fig.~\ref{fig:monte_chi} shows the equivalent results for the Bayesian test statistic, $\chi^n$.  Also shown on the plots are dotted lines indicating the threshold test levels for the false alarm and missed detection probabilities described above.  In both cases, there were 40 missed detections for the 50,000 runs with planets.  For PSF fitting there were 3 false alarms and in the Bayesian approach there were 2.  These plots confirm the analytical results and also justify the Gaussian assumption.
 
\begin{figure}
 \plotone{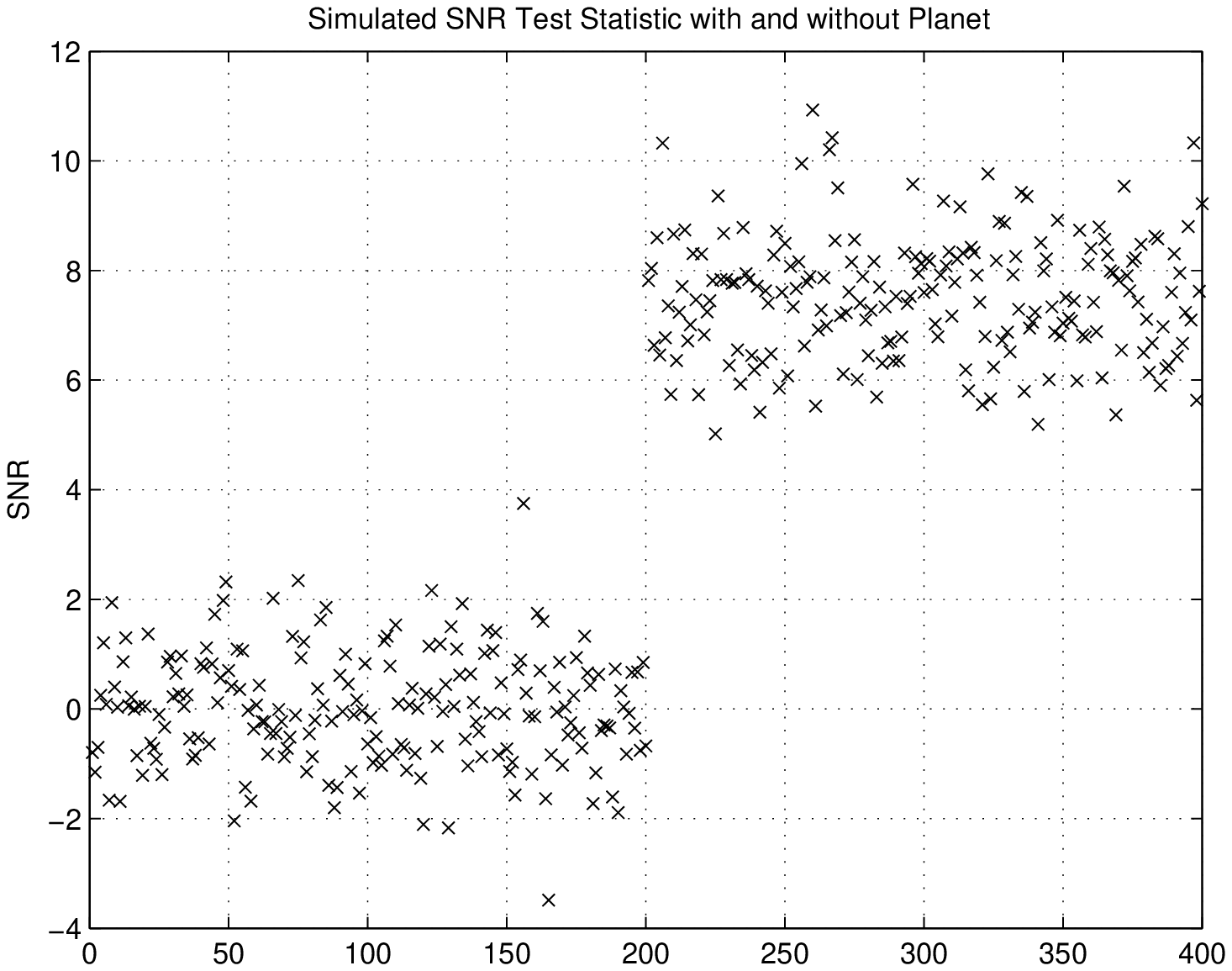}
 \caption{Samples of the simulated PSF fitting test statistic with and without a planet for the selected integration time. }
  \label{fig:snr_sim}
 \end{figure}

 Figure~\ref{fig:snr_sim} shows 200 samples each of the PSF fitting test statistic with and without a planet.  This illustrates how the test statistic indicates the presence of a planet by  exceeding the zero mean scatter by a threshold amount.
 
 \section{Final Remarks}

In this paper we developed a careful model of the coronagraph measurement and used it to define two different detection approaches, matched filtering and Bayesian.  For either approach, we developed a simple expression for the needed integration time to achieve the desired probability of success.  These integration time results can now be utilized for mission completeness studies.

The story is far from complete, however.  Critical to these studies is an understanding of the expected background and the best possible estimates of $Q$.  Additionally, we have ignored the question of wavefront error and control.  In this simplified model, the background was assumed known and uniform.  In reality, the background is going to be a mix of exozodiacal light and residual wavefront error.  The speckle background due to uncorrected wavefront error will not be uniform and can potentially confuse the planet detection algorithm.  While the planet and background can be simultaneously estimated, it complicates the algorithm and lengthens the integration time.  Also, the additional flexibility of multi-wavelength measurements has not been explored.  Finally, the sensitivity questions of mismatches in assumed parameters and PSF alignment on pixels has not yet been studied.  These issues are all currently being addressed and will be presented in a future paper.  Nevertheless, the fundamental model and technique presented here provides an important benchmark to measure the potential search capabilities of TPF.

\bibliographystyle{unsrtnat}
\bibliography{../../TPF/Planet_Finding_NRA/Phase_II/FinalReport}

\appendix
\section{Photometry via PSF Fitting}
 
 In this appendix we briefly summarize the derivation of Eq.~\ref{eq:Cp_estimate}.  This is a slightly expanded treatment of the same problem in \cite{burrows03}.   It is also demonstrates the overdetermined, MLE case in \cite{lyon97}.  As presented there, the problem is to estimate the irradiance, or photon count, of a signal corresponding to a point source in a known, uniform background.  The image is given by the measurement $I_{ij} = B + AP_{ij} + N_{ij}$, where $B$ is the assumed known background, $A$ is the source irradiance, $P_{ij}$ is the point spread function spread over the pixels $(i,j)$, and $N_{ij}$ are noise photons in each pixel $(i,j)$ with zero mean and variance $\sigma^2_{ij}$.  The noise is assumed independent and identically distributed across the pixels, so that $E\{N_{ij} N_{kl}\} = \sigma^2_{ij} \delta_{ik} \delta_{jl}$. What makes this problem more complicated than the standard linear least squares estimate is that the noise is Poisson distributed, so the variance at pixel $(i,j)$ is given by $\sigma^2_{ij} = B + AP_{ij} $, (where we are ignoring the readout noise).  
 
 \subsection{Linear Least Squares Estimate}

Since the background $B$ is known, the weighted least squares estimate is found by minimizing the following cost with respect to A,
\begin{equation}
J = \frac{1}{2} \sum_{ij} W_{ij} (I_{ij} - B - AP_{ij})^2
\end{equation}
where the $W_{ij}$ are some to be determined weights.

This results in the optimal estimate,
\begin{equation}
\label{photo_ls}
\hat{A} = \frac{\sum_{ij} (I_{ij} - B)W_{ij} P_{ij}}{\sum_{ij} W_{ij} P^2_{ij}}
\end{equation}
The optimal weights are found by forming the variance of the estimate and minimizing with respect to $W_{ij}$ (minimum variance estimation),
\begin{equation}
\sigma_A^2 = \frac{\sum_{ij} \sigma^2_{ij} P^2_{ij} W^2_{ij}}{(\sum_{ij} W_{ij}P^2_{ij})^2}
\end{equation}
The result is that $W_{ij} \propto 1/\sigma^2_{ij}$.  That is, as is well known, the optimal weights are given by the noise variance.  This implies that the cost function to be minimized is given by,
\begin{equation}
\label{eq:J_nonlin}
J = \frac{1}{2} \sum_{ij} \frac{(I_{ij} - B - AP_{ij})^2}{AP_{ij} + B}
\end{equation}
This equation has theoretical appeal since for large photon counts, the Poisson noise approaches a Gaussian distribution and this is related to the maximum likelihood estimate.  Note, however, that $A$ appears nonlinearly in the cost and an exact maximum likelihood solution requires a more complex calculation.  We address that further below.  

It is important to note that there is a bit of circular reasoning here.  The expression for $\hat{A}$ in Eq.~\ref{photo_ls} was found by minimizing the cost with respect to $A$ assuming the weights were independent of $A$.   In the standard problem formulation, this is usually the case and the resulting estimate in Eq.~\ref{photo_ls} is optimal.  In  this case, however, this result was used to find that the optimal weights were proportional to the noise variance, which for Poisson noise is a function of $A$!  This puts into doubt the optimality of the resulting expression.  However, the case of most interest here is the background limited case where $B \ggg AP_{ij}$.  Here, the weights become constant and thus cancel out of the estimate.  The resulting  estimate is then given by,
\begin{equation}
\label{photo_ls_final}
\hat{A} = \frac{\sum_{ij} (I_{ij} - B) P_{ij}}{\sum_{ij} P^2_{ij}}
\end{equation}
and the variance of the error is given by,
\begin{equation}
\sigma_A^2 = \frac{\sum_{ij} P^2_{ij} \sigma^2_{ij}}{(\sum_{ij}P^2_{ij})^2} = \frac{B}{\sum_{ij} P^2_{ij}}
\end{equation}
where, in this background limited case, $\sigma^2_{ij} \equiv B$.

A typical measure of the estimate quality is the signal-to-noise ratio given by  dividing $A$ by the standard deviation of its estimate and squaring, resulting in, for the background limited case,
\begin{equation}
\label{SNR}
\left ( \frac{A}{\sigma_A} \right )^2 = \frac{A^2 \sum_{ij} P^2_{ij}}{B}
\end{equation}
Despite being non-optimal for the case where the background and planet are comparable, Eq.~\ref{photo_ls_final} has the appeal of being unbiased; it is therefore often used as the estimate for the general case as well.  It is tempting to attempt a nonlinear least squares solution of the cost in Eq.~\ref{eq:J_nonlin}, hoping that, though nonoptimal (the weights don't represent the minimum variance because the dependence on $A$ was ignored), that it will still provide an improved $SNR$. Unfortunately, this is not the case.  Not only is the problem much more difficult, with no simple closed form solution, the resulting estimate is biased and the variance is larger.

\subsection{BLUE estimate}

In his appendix, \cite{burrows03} used the BLUE ({\em B}est {\em L}inear {\em U}nbiased {\em E}stimate) approach  rather than least squares minimization.  In fact, this is sensible as it avoids  some of the circular reasoning problems above.  We reproduce that here and show that the same signal-to-noise ratio results.

The most general linear, unbiased estimate of the irradiance is given by,
\begin{equation}
\label{photo_blue}
\hat{A} = \frac{\sum(I_{ij} - B)W_{ij}}{\sum W_{ij} P_{ij}}
\end{equation}
It is straightforward to show that this is an unbiased estimate for any $W_{ij}$ as long as $N_{ij}$ is zero mean.  The variance of this estimate is found via,
\begin{eqnarray}
\label{blue_variance}
E\{(A-\hat{A})^2\} &= &\frac{\sum_{ij} \sum_{kl} \delta_{ik}\delta_{jl} \sigma_i^2 W_{ij} W_{kl}}{(\sum_{ij} W_{ij} P_{ij})^2} \nonumber \\
\sigma_A^2 & = & \frac{\sum_{ij} \sigma_{ij}^2 W_{ij}^2}{(\sum_{ij} W_{ij} P_{ij})^2}
\end{eqnarray}
This is the same expression as in Eq. (D.1) of \cite{burrows03}.  Following the same procedure as above, the optimal weights are found by maximizing the signal-to-noise ratio or, equivalently, minimizing the variance.  The resulting expression for the optimal weights is,
\begin{equation}
W_{ij} = \frac{P_{ij}}{\sigma^2_{ij}} \left ( \frac{\sum_{ij} \sigma^2_{ij} W^2_{ij}}{\sum_{ij} W_{ij} P_{ij}} \right )
\end{equation}
This shows that the weights are proportional to $P_{ij}/\sigma^2_{ij}$, as the term in parantheses is independent of $(i,j)$.  Since any constant term cancels out of the estimate, the weights can be set equal to $P_{ij}/\sigma^2_{ij}$ without loss of generality,
\begin{equation}
W_{ij} = \frac{P_{ij}}{AP_{ij} + B}
\end{equation}
The circularity problem  of Poisson processes shows up here when this expression for the weights is substituted into the BLUE estimate in Eq.~\ref{photo_blue}, as the expression for $\hat{A}$ depends upon $A$ itself.  This is because the exact problem is inherently nonlinear; the   linear estimate is only unbiased if the weights are forced to be independent.  For an exact optimal estimate, one must turn to maximum likelihood, or other, approaches.  However, again assuming a background limited signal, where $AP_{ij} \ll B$, the weights can be taken to be $W_{ij} = P_{ij}$.  Substituting this into Eq.~\ref{photo_blue} results in the estimate equation,
\begin{equation}
\label{photo_blue_final}
\hat{A} = \frac{\sum(I_{ij} - B)P_{ij}}{\sum P^2_{ij}}
\end{equation}
This is exactly the same as Eq.~\ref{photo_ls_final} for the least squares estimate and what we use for the PSF fitting detection in Eq.~\ref{eq:Cp_estimate}.

As before, the expression for the weights, $W_{ij} = P_{ij}$, can be substituted into the variance expression (Eq.~\ref{blue_variance} above) to form a signal-to-noise ratio equation,
\begin{equation}
\label{eq:snr_blue}
\left ( \frac{A}{\sigma_A} \right )^2 = \frac{A^2 \sum_{ij} P^2_{ij}}{B}
\end{equation}

It is interesting to look at the opposite case where the background is negligible.  In that case, $W_{ij} \cong 1/A$, a constant that cancels out of Eq.~\ref{photo_blue}.  This is equivalent to the least squares solution in the previous subsection weighted only by the PSF.  This is a very common photometric approach to PSF fitting.  The  estimate simply becomes,
\begin{equation}
\label{eq:BLUE_nobackground}
\hat{A} = \frac{\sum I_{ij}}{\sum P_{ij}}
\end{equation}
with variance,
\begin{equation}
\sigma^2_A = \frac{A}{\sum P_{ij}}
\end{equation}
This is an unbiased estimate with SNR of $\sqrt{A \sum P_{ij}}$.

It is important to note that the estimator in Eq.~\ref{photo_blue_final} is still unbiased even if the background and signal are comparable; it is just no longer minimum variance.   It is for that reason that it is used as the estimator for the detection problem.  It interesting, however, to compute the variance and resulting signal-to-noise ratio of the estimator in Eq.~\ref{photo_blue_final} but with the full noise variance.  The  expression for the estimate variance becomes:
\begin{equation}
\sigma_A^2 = \frac{A\sum_{ij} P^3_{ij}}{(\sum_{ij}P^2_{ij})^2} + \frac{B}{\sum_{ij} P^2_{ij}}
\end{equation}
resulting in an expression for the signal-to-noise ratio including the point source photon noise,
\begin{equation}
\label{eq:snr_general}
\left ( \frac{A}{\sigma_A} \right )^2 =\frac{A^2 (\sum_{ij} P^2_{ij})^2}{A\sum_{ij}P^3_{ij} + B\sum_{ij} P^2_{ij}}
\end{equation}

Because of the Poisson nature of the signal and the resulting dependence of the noise variance on the signal itself, the linear, unbiased estimator is no longer optimal.  Nevertheless, the linearity of the equation makes performing a detection and finding integration time estimates  straightforward.  It is, therefore, the most common approach to PSF fitting.  It is, however, appealing to investigate the possibility of a maximum likelihood estimator that accounts for this dependence.  We summarize such an approach next.

\subsection{Poisson Maximum Likelihood Estimation}

Since the photons are known to arrive via a Poisson process, it is sensible to turn to maximum likelihood estimation  under the Poisson distribution in an attempt to solve the optimality problem.\footnote{It is tempting to first search for a ML estimator assuming a Gaussian distribution of the noise, particularly since for large photon counts the Poisson distribution is close to Gaussian.  However, Gaussian maximum likelihood in this case is difficult, highly nonlinear, and results in a biased estimate with larger error than the approximate linear cases.}  For simplicity, we start by examining the zero background case.  The Poisson probability distribution at pixel  (i,j) is given by,
\begin{equation}
p_{ij} = e^{-AP_{ij}} \frac{\left ( A P_{ij} \right )^{I_{ij}}}{I_{ij}!}
\end{equation}
The  total measurement probability across all the pixels is the product of the individual probabilities, as the measurements at each pixel are independent:
\begin{equation}
p = e^{-\sum A P_{ij}} \frac{ \prod \left ( A P_{ij} \right )^{I_{ij}}}{\prod I_{ij}!}
\end{equation}
Taking the logarithm gives the log-likelihood function,
\begin{equation}
L = -\sum A P_{ij} + \sum I_{ij} \ln(A)
\end{equation}
where we have dropped constant terms independent of $A$.  Taking the derivative and setting equal  to zero gives  the maximum likelihood estimate,
\begin{equation}
\hat{A} = \frac{\sum I_{ij}}{\sum P_{ij}}
\end{equation}
This is identical  to the BLUE estimate with  zero background in Eq.~\ref{eq:BLUE_nobackground} and the PSF weighted linear least squares.  For signals with little to no background, this is the  most common optimal estimate of the intensity.

For the Poisson maximum likelihood with background, we form the total probability density function,
\begin{equation}
p = e^{-\sum A P_{ij} + NB } \frac{ \prod \left ( A P_{ij} + B  \right )^{I_{ij}}}{\prod I_{ij}!}
\end{equation}
where $N$ is the number of pixels used in the estimate sums.    We again form the log-likelihood function,
\begin{equation}
L = -\sum A P_{ij} + \sum I_{ij} \ln(AP_{ij} + B)
\end{equation}
and take the derivative and set equal to zero to find the equation that $\hat{A}$ must satisfy,
\begin{equation}
\label{eq:poisson_ml}
\sum \frac{I_{ij} P_{ij}}{\hat{A}P_{ij} + B} = \sum P_{ij}
\end{equation}

For the general case of comparable background and signal, there is no closed form solution to this  expression for $\hat{A}$.  One must resort to numerics.  However, taking the expected value of both sides does show that the estimate is unbiased.  At the cost of a loss of linearity, this is the best  PSF fitting estimate for $A$, though it is difficult to find an expression for the signal-to-noise ratio that can be used to determine integration times.  Monte-Carlo approaches are the most convenient.

For the background limited case ($B \gg AP_{ij}$) it is possible to  find an approximate expression for $\hat{A}$ by expanding Eq.~\ref{eq:poisson_ml},
\begin{equation}
\hat{A} \cong \frac{B}{\sum I_{ij} P_{ij}^2} \left ( \sum I_{ij} P_{ij} - B \sum P_{ij} \right )
\end{equation}
Again, the assumption is that the background is known.  Because of the first order expansion, this estimate has a bias of $\mathcal{O}(1/B)$, small for the background dominated case.  Because this ML expression does not lead to useful expressions of the integration time, we continue to use the (perhaps) more conservative linear approach.

\end{document}